\documentclass[aps,prd,twocolumn]{revtex4-2}

\usepackage[colorlinks]{hyperref}
\usepackage{amssymb,amsmath}
\usepackage{dsfont}
\usepackage{graphicx}

\begin{document}

\title{A multishell solution in the Skyrme model}

\author{V. V. Flambaum}
\email{v.flambaum@unsw.edu.au}
\affiliation{School of Physics, University of New South Wales, Sydney 2052, Australia}

\author{I. B. Samsonov}
\email{igor.samsonov@unsw.edu.au}
\affiliation{School of Physics, University of New South Wales, Sydney 2052, Australia}

\begin{abstract}
We consider multishell configurations in the Skyrme model within the rational map ansatz. We show that equations for the Skyrme field are linearized in the limit of large number of shells, thus allowing for a simple analytic solution. Although this solution is approximate, it provides an accurate description of multishell configurations in the Skyrme model in the region where the Skyrme field is large, $F\gg1$. We use this solution to calculate the mass and the root mean square radius of multishell skyrmion configurations. In particular, for solutions with one unit of baryon charge per shell (the ``hedgehog'' solution) the mass scales as $M\propto B^2$, and its rms radius scales as $B^{1/2}$ with the baryon charge $B$. This scaling for the mass can be reduced to $M\propto B^{4/3}$ in the model with many units of baryon charge per shell. Although this solution is unstable against decays into single-shell or single-skyrmion  configurations, it may be useful for modelling skyrmion stars or compact composite objects in some models of dark matter if the decay of such configurations is prevented by some mechanism.
\end{abstract}

\maketitle

\section{Introduction}

The Skyrme model \cite{Skyrme} is a non-linear sigma model of pseudoscalar meson fields with quartic selfinteraction. This model allows for static compact solutions with finite energy which may be considered as baryons \cite{Adkins83}. The baryon charge $B$ is naturally identified with the topological charge number in the Skyrme model. The solution with $B=1$ provides a satisfactory description of a nucleon, see, e.g., Ref.~\cite{Review1} for a review.

Solutions with $B>1$ are usually considered as systems of interacting nucleons. In Ref.~\cite{Rho83}, spherically symmetric skyrmions with $B=2$ and $B=3$ were studied numerically. It was shown that these solutions are unstable, and thus should decay into spatially separated solutions with $B=1$. 

Recent developments in the Skyrme model were devoted to constructing multi-$B$ solutions with reduced symmetry which could model atomic nuclei. In Ref.~\cite{Kopeliovich,Manton,Verb}, an axially symmetric solution in the Skyrme model was found for $B=2$. In Refs.~\cite{Braaten,Sutcliffe97,Houghton1997}, solitonic configurations with $B\leq9$ were constricted. Notably, crystal-like solutions with $B\to\infty$ were found in Refs.~\cite{Klebanov,crystal}. 

The solutions of the Skyrme model proposed in Refs.~\cite{Sutcliffe97,Houghton1997} within the rational map ansatz have a shell-like structure. The number of units of baryon charge per shell is equal to the degree of the rational map. In Refs.~\cite{piette2007,Nelmes2011} it was conjectured that such multishell solutions with a large number of units of baryon charge per shell may describe skyrmion stars, analogs of neutron stars. The authors of these works managed to study numerically only large-$B$ solutions with two shells and approached multishell configurations semianalytically using ramp-profile ansatz that may give overestimated value of the mass.

In this note, we focus on multishell solutions in the Skyrme model which have a large number of shells and arbitrary number of units of baryon charge per shell. In Sec.~\ref{SecSolution}, we show that equations for stationary skyrmion configurations within the rational map ansatz are linearized when the number of shells goes to infinity. As a result, we find a simple analytic solution describing multishell skyrmion configurations which may be used for studying skyrmion stars or compact composite objects playing role of dark matter particles in some dark matter models \cite{QN1,QN2,QN3,QN4}. We show that the mass of this solution scales as $B^2$ while its root mean square radius grows as $B^{1/2}$. Thus, the obtained solution is unstable against decays into single-shell or single-skyrmion configurations, and it may be realized only under external conditions preventing them from the decay. In Sec.~\ref{SecSummary} we discuss the results and speculate that adding the pion mass term to the Skyrme model may help stabilize multishell solutions.

In this paper, natural units with $\hbar=c=1$ are used.

%%%%%%%%%%%%%%%%%%%%%%%%%%%%%%%%%%%%%%%%%%%%%%%%%%

\section{The Skyrme model}

In this section, we provide a background information about the Skyrme model, which is necessary for completeness of our presentation in this paper.

At low energy, baryons and mesons represent the effective degrees of freedom emerging from QCD. The mesons are dominated by pseudoscalar pion fields $\vec\pi$ which are described by a non-linear sigma model based on the $SU(2)_L\times SU(2)_R$ group. These fields constitute the following matrix $U(x) = f_\pi^{-1}( \mathds{1} \sigma + i\vec\tau\cdot \vec\pi)\in SU(2)$, where $f_\pi = 93$~MeV is the pion decay constant and $\sigma$ is the scalar meson with the constraint $\sigma^2 + \vec\pi^2 =f_\pi^2$. The field $U(x)$ transforms in the $(\frac12,\frac12)$ representation of $SU(2)_L\times SU(2)_R$: $U(x)\to A U B^{-1}$, $A\in SU(2)_L$ and $B\in SU(2)_R$. 

At any fixed time, $U(x)$ maps the space $\mathds{R}^3$ onto the group manifold $S^3\simeq SU(2)$. Since the pion fields vanish at spatial infinity, the space $\mathds{R}^3$ may be compactified to the sphere $S^3$, and the above map turns into $U(x):S^3\to S^3$. This map is charaterized by a topological charge density $B^0$ arising as a 0-component of a conserved 4-current 
\begin{equation}
\label{Bmu}
B^\mu = \frac{\varepsilon^{\mu\nu\rho\sigma}}{24\pi^2}
\text{tr}[(U^\dag \partial_\nu U)(U^\dag \partial_\rho U)(U^\dag \partial_\sigma U)]\,,
\end{equation}
with $\varepsilon_{0123}=-\varepsilon^{0123}=1$.

The Skyrme model is a non-linear sigma-model of the pion fields $\vec\pi$ with the Lagrangian
\begin{equation}
\label{L}
    {\cal L} = \frac14 f_\pi^2 \text{tr}(\partial_\mu U\partial^\mu U^\dag) + \frac1{32e^2}\text{tr}
    [(\partial_\mu U)U^\dag , (\partial_\nu U)U^\dag]^2\,,
\end{equation}
where $e$ is a dimensionless coupling constant. A skymion is a static solution in this model of the form
\begin{equation}
\label{ansatz}
U(x) = \exp[iF(r) \vec\tau\cdot\vec n]\,,
\end{equation}
where $\vec n$ is a unit vector and $F(r)$ is a spherically symmetric function with boundary conditions
\begin{equation}
\label{BC}
    F(r)|_{r=0}= b\pi\,,\qquad
    F(r)|_{r\to\infty}\to0\,.
\end{equation} Here $b$ is an integer. 

In the original Skyrme model \cite{Skyrme}, the unit vector $\vec n$ is chosen in a spherically symmetric way, $\vec n = \vec r/r$. This case is suitable for studying solution with $b=1$, which is usually interpreted as a models of nucleon. Solutions with $b>1$ were shown to be unstable \cite{Rho83}.

More generally, the unit vector in Eq.~(\ref{ansatz}) may be chosen in the form \cite{Sutcliffe97}
\begin{equation}
\label{Ransatz}
    \vec n = \frac{1}{1+|R|^2}(2\text{\,Re\,}(R),2\text{\,Im\,}(R),1-|R|^2)\,,
\end{equation}
where $R$ is a rational function on the complex plane with coordinates $z$ related to the spherical angles $\theta$ and $\phi$ of $\mathds{R}^3$ through the stereographic projection $z = \tan(\theta/2)e^{i\phi}$. It was shown in Ref.~\cite{Sutcliffe97} that the total baryon charge $B$ within this ansatz is given by 
\begin{equation}
\label{kb}
B = kb\,,
\end{equation}
where $b$ is an integer specifying the boundary condition (\ref{BC}) and $k$ is the degree of the rational map $R$. This ansatz allows for stable multi-$B$ solutions which satisfactory describe some atomic nuclei \cite{Sutcliffe2009,Manton2013}.

Substituting the ansatz (\ref{Ransatz}) into the Lagrangian (\ref{L}) one finds the expression for the skyrmion mass
\begin{equation}
\begin{aligned}
\label{M}
    M =& 2\pi f_\pi^2 \int_0^\infty \left[r^2(F')^2 + 2k\sin^2 F\right] dr \\ 
    &+\frac{2\pi}{e^2} \int_0^\infty \left[
        \frac{{\cal I}(k)\sin^4 F}{r^2} + 2k(F')^2\sin^2F
    \right]dr\,,
\end{aligned}
\end{equation}
with $F'= dF/dr$ and
\begin{equation}
    {\cal I }(k) = \frac1{4\pi}\int \left(\frac{1+|z|^2}{1+|R|^2}\left|\frac{dR}{dz} \right| \right)^4
    \frac{2idzd\bar z}{(1+|z|^2)^2}\,.
\end{equation}
Numerical values of this function for various $k$ were studied in Ref.~\cite{Sutcliffe97}. In particular, ${\cal I}=1$ for $k=1$, and 
\begin{equation}
{\cal I}(k)|_{k\to\infty} \to {\cal I}_\infty k^2\,,\qquad
{\cal I}_\infty \approx 1.28\,.
\end{equation}

Varying the expression (\ref{M}) we obtain the equation for the function $F$:
\begin{equation}
\label{EOM}
\begin{aligned}
    & \left[(r^2 +2k r_\text{sk}^2 \sin^2 F)F'\right]' = \sin 2F \bigg[
    k  \\ & +kr_\text{sk}^2 (F')^2 
    +{\cal I}(k)\frac{r_\text{sk}^2}{r^2}\sin^2 F \bigg]\,,
\end{aligned}
\end{equation}
where $r_\text{sk} = \frac1{ef_\pi}$.
 
The baryon charge density (\ref{Bmu}) acquires a simple expression within the ansatz (\ref{ansatz}):
\begin{equation}
\label{rhoB}
    \rho_B \equiv B^0 = -\frac{k}{2\pi^2} \frac{\sin^2 F}{r^2}F'\,.
\end{equation}
By construction, integrating this baryon charge density one finds the baryon charge number,
\begin{equation}
\label{DensityNormalization}
    4\pi k\int_0^\infty \rho_B r^2\,dr = kb= B\in \mathds{Z}\,.
\end{equation}
In this paper we focus on the case $b\gg1$ with arbitrary $k$.

%%%%%%%%%%%%%%%%%%%%%%%%%%%%%%%%
\label{SecSolution}

\section{A large-B solution}

Equation (\ref{EOM}) is non-linear, and its exact analytic solutions are not known. Numerical solutions of this equations were studied in Refs.~\cite{Adkins83,Rho83} for $B=1,2,3$. In this section we construct an explicit analytic solution of this equation for $B\gg1$. More precisely, we consider a multishell solution with a large number of shells and arbitrary number of units of baryon charge per shell. 

The boundary condition (\ref{BC}) shows that near the origin the solution is large, $F\gg1$ if $b\gg1$, and it drops rapidly with the distance. Thus, in this region the functions $\sin 2F$ and $\sin^2F$ are rapidly oscillating with the following average values
\begin{equation}
\label{average}
    \langle \sin 2F \rangle =0\,,\qquad
    \langle \sin^2 F \rangle =\frac12\,.
\end{equation}
With these values of the trigonometric functions, Eq.~(\ref{EOM}) is linearized,
\begin{equation}
\label{EOM-short}
    \left[(r^2 +kr_\text{sk}^2) F'\right]'  =0\,.
\end{equation}
A solution of this equation subject to the boundary condition (\ref{BC}) reads
\begin{equation}
\label{solution}
    F(r) = 2b \,\text{arccot}\frac{r}{\sqrt{k} r_\text{sk}}\,.
\end{equation}

Note that a similar solution was found in a self-dual modified Skyrme model \cite{Ferreira}.

\subsection{Mass of the skyrmion configuration}

Making use of Eq.~(\ref{M}) it is straightforward to find the mass of the skyrmion configuration (\ref{solution}):
\begin{equation}
\label{Mhedgehog}
    M = 4\pi^2 f_\pi e^{-1}\left[ b^2\sqrt{k} +
    bk\sqrt{k}(1+\tfrac14{\cal I}_\infty)\right].
\end{equation}
Recalling that the baryon charge is $B=bk$, we find that the mass per unit baryon charge scales as 
\begin{equation}
\label{MB}
    \frac{M}{B} = 4\pi^2 \frac{f_\pi}{e}
   \left[ \frac{b}{\sqrt{k}} + \sqrt{k} \left( 1+\frac{{\cal I}_\infty}{4}\right)\right]. 
\end{equation}

The expression (\ref{MB}) allows us to study the scaling of the ratio $M/B$ for large baryon charge $B$. Recall that $b$ is the number of shell and $k$ is the number of baryon charge units per shell. Consider first the case $k=1$, which corresponds to the original ``hedgehog'' solution in the Skyrme model. As is seen from Eq.~(\ref{Mhedgehog}), $M\sim B^2$ in this case. This scaling was first established in Ref.~\cite{Bogomolnyi}. With the use of the explicit analytic solution (\ref{solution}) we find exact coefficients in this relation.

Consider now the case when both $b\gg1$ and $k\gg1$. It is natural to assume $k\propto b^2$, meaning that the number of skyrmions per shell is proportional to the area of a sphere of radius $b$. In this case Eq.~(\ref{MB}) shows that $M/B\propto \sqrt{k}\propto B^{1/3}$. Thus, solutions in the Skyrme model with multiple shells are unstable against decays into single-shell or single-skyrmion configurations.

\subsection{Root mean square radius}

Note that the baryon charge density (\ref{rhoB}) is normalized as in Eq.~(\ref{DensityNormalization}). Therefore, the mean radius squared is defined as
\begin{equation}
    \langle r^2 \rangle = B^{-1} \int r^2 \rho_B(r) d^3r\,.
\end{equation}
Substituting the solution (\ref{solution}) into this equation we find 
\begin{equation}
    \langle r^2 \rangle = 4B r_\text{sk}^2(1-\frac1{4b})\,.
\end{equation}
The last term may be neglected in the large-$b$ limit. Thus, for large $B$, we find 
\begin{equation}
    r_\text{rms} \equiv \sqrt{\langle r^2 \rangle} = 2\sqrt{B} r_\text{sk}\,.
\end{equation}
As a result, the characteristic size of this solution grows with $B$ faster than that of the nuclear matter. For the latter, we recall that the nuclear charge radius scales with the atomic number as $A^{1/3}$.

\subsection{Asymptotic behaviour}

The solution (\ref{solution}) has the following asymptotic behaviour:
\begin{equation}
    F(r)|_{r\to\infty} \to 2\frac{B}{\sqrt{k}} \frac{r_\text{sk}}{r}\,.
\end{equation}
This behaviour is, however, unphysical, because at large $r$ Eq.~(\ref{EOM}) reduces to $r^2 F'' + 2r F' - 2kF =0$, that dictates $F|_{r\to\infty} \propto r^{-\frac12 -\frac12\sqrt{8k+1}}$. Thus, the solution (\ref{solution}) is applicable only in the interval $0\leq r < R$, in which $F(r)\gg1$. Here $R$ is a cut-off radius for the solution (\ref{solution}).

It is convenient to define the cut-off radius $R$ such that $F(R) = 2$. In this case, for $b\gg1$,
\begin{equation}
\label{Rcutoff}
    R = \sqrt{k} r_\text{sk} \cot \frac1{b} \approx \frac{B}{\sqrt{k}} r_\text{sk}\,.
\end{equation}
For $r> R$, the solution (\ref{solution}) needs to be corrected.

\subsection{Approximate solution with corrected asymptotic behaviour}

It is possible to construct a function $F_\text{approx}(r)$ such that $F_\text{approx}(r)\approx F(r)$ for $0<r<R$, and  it possesses the correct asymptotics $F_\text{approx}(r)|_{r\to\infty}\propto r^{-\frac12 -\frac12\sqrt{8k+1}}$. Of course, such function is just an approximate solution of the model, and it may be constructed in different ways. In particular, it is convenient to use the following function:
\begin{equation}
\label{Fapprox}
    F_\text{approx}(r) = 2b\left(\frac{R}{r+R}\right)^{-\frac12
+\frac12\sqrt{8k+1}} \text{arccot}\frac{r}{\sqrt{k}r_\text{sk}}\,,
\end{equation}
which reduces to (\ref{solution}) for $r\ll R$, but asymptotically it behaves as 
\begin{equation}
    F_\text{approx}(r)|_{r\to\infty} \to 2\left( 
    \frac{B}{\sqrt{k}} \frac{r_\text{sk}}{r}
    \right)^{\frac12+\frac12\sqrt{8k+1}}\,.
\end{equation}

\begin{figure*}
    \centering
    \begin{tabular}{cc}
    \includegraphics[width=8cm]{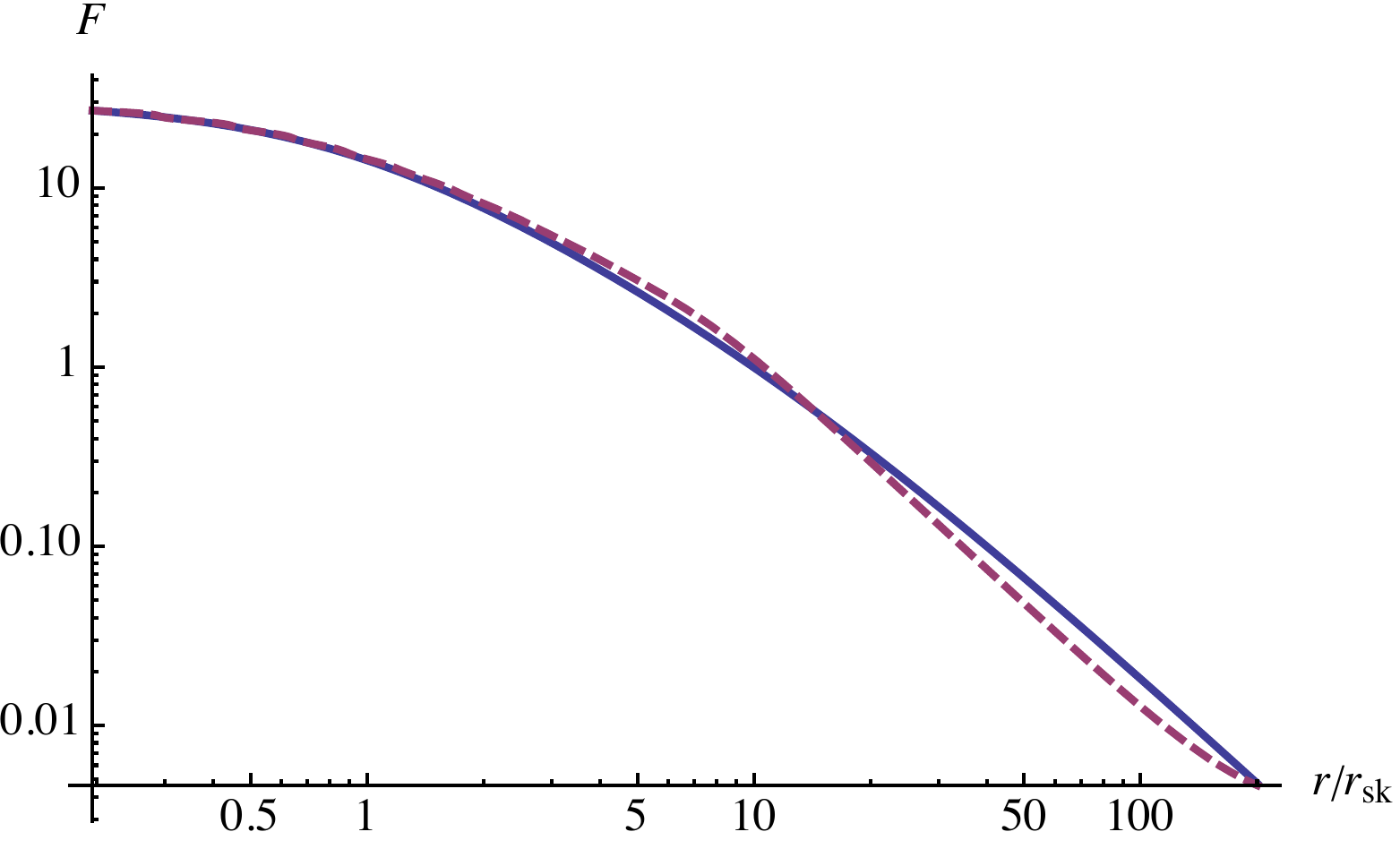} &
        \includegraphics[width=8cm]{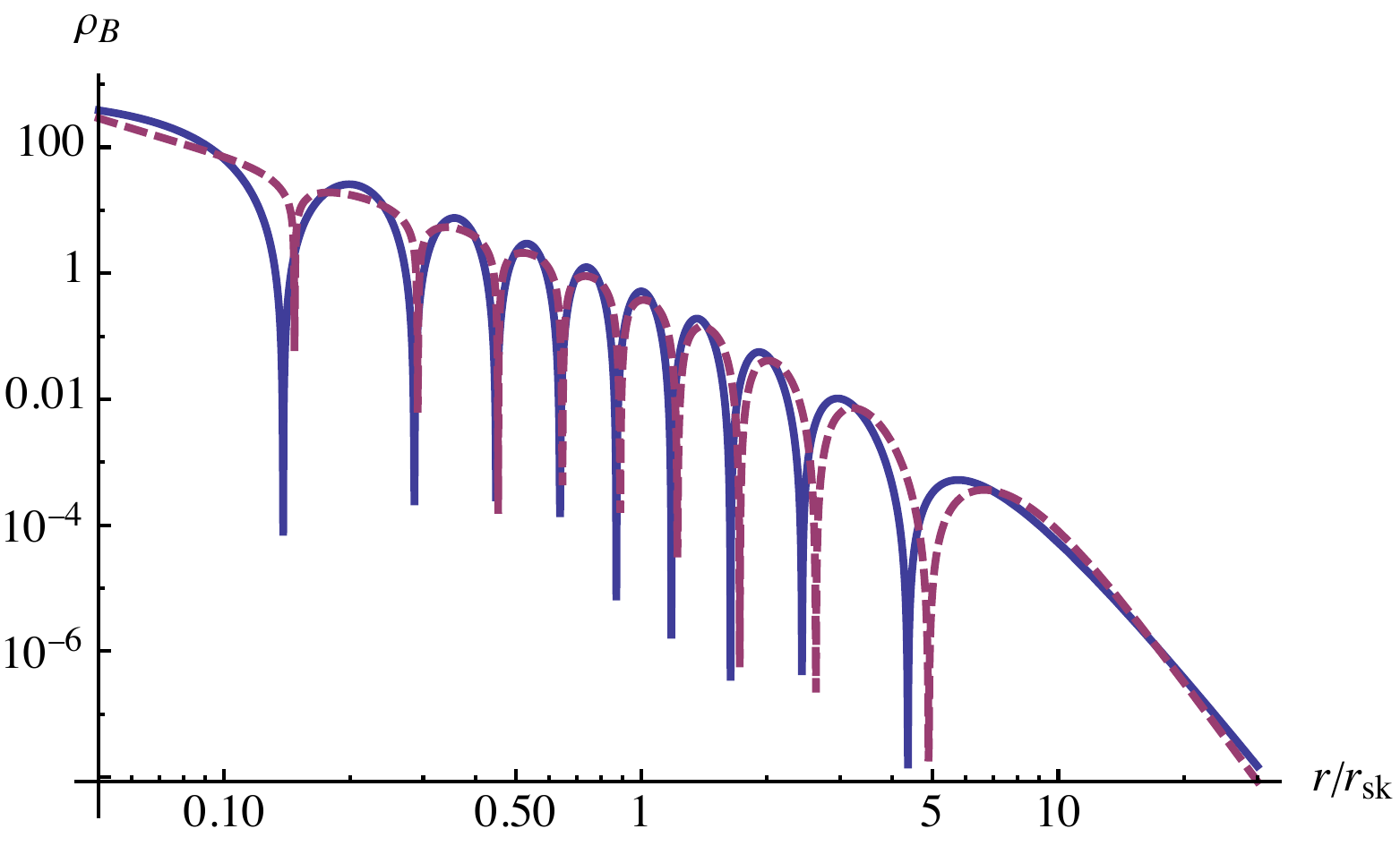} \\
        a & b
    \end{tabular}
    \caption{ (a) Comparison of the approximate solution (\ref{Fapprox}) (solid blue curve) with a numerical solution of Eq.~(\ref{EOM}) (dashed red curve). (b) Plot of the baryon charge density (\ref{rhoB}) calculated for the approximate solution (\ref{Fapprox}) (solid blue curve), and for exact numerical solution of Eq.~(\ref{EOM}) given by red dashed curve. Both figures are produced for $b=10$ and $k=1$ that corresponds to ten shells with one unit of baryon charge per shell.}
    \label{fig1}
\end{figure*}

The approximate solution (\ref{Fapprox}) is compared with exact numerical solution of Eq.~(\ref{EOM}) in Fig.~\ref{fig1}a. It shows a good agreement with the numerical solution both at small and large distance. We plotted also the baryon charge density (\ref{rhoB}) for the solution (\ref{Fapprox}). As is seen from Fig.~\ref{fig1}b, the multi-$B$ skyrmion has an onion shell-like structure with $b$ shells and $k$ units of the baryon charge in each shell.

%%%%%%%%%%%%%%%%%%%%%%%%%%%%%%%%%%%%%%%%%%%%%%%%%%%

\section{Summary and discussion}
\label{SecSummary}

In this note, we constructed an explicit analytic multishell solution (\ref{solution}) in the Skyrme model. This solution appears due to the observation that equation (\ref{EOM}) for the Skyrme field $F(r)$ is linearized for $F\gg1$. Thus, this analytic solution provides an accurate description of multishell configurations in the Skyrme model in the interval $0<r<R$, in which $F(r)\gg1$. The cut-off parameter $R$ is given in Eq.~(\ref{Rcutoff}). For $r>R$, the solution (\ref{solution}) should be modified, because it falls off with distance slower than asymptotic solutions of Eq.~(\ref{EOM}). One possible long-distance modification of the solution (\ref{solution}) is proposed in Eq.~(\ref{Fapprox}).

The analytic solution (\ref{solution}) allows us to calculate exactly the mass of multishell configurations in the Skyrme model, see Eq.~(\ref{Mhedgehog}). In a particular case of one baryon charge per shell, $k=1$, the mass of such solution scales as $M\propto B^2$ with the baryon charge $B$. This scaling was first found in Ref.~\cite{Bogomolnyi} whereas in this paper we determine exact coefficient in this relation. In the case of arbitrary number of units of baryon charge per shell we show that the mass per unit baryon charge grows at least as $\sqrt{k}$, and, thus, any multishell solution is unstable against decays into single-shell or single-skyrmion configurations.  

\subsection{Adding pion mass term}

As is noted above, the mass of the multi-$B$ skyrmion configuration (\ref{solution}) grows quadratically with the baryon charge, $M\propto B^2$, for $k=1$. This behaviour is characteristic for a set of $B$ particles with a long-range two-body interaction. Therefore, it is natural to expect that this behaviour may change in a Skyrme model with massive pion fields. 

In Ref.~\cite{Kopeliovich} it was shown that adding the pion mass term to the Lagrangian reduces the mass-per-baryon ratio in a multi-$B$ solution within the rational map ansatz. Moreover, for some choices of the mass term, the mass-per-baryon ratio tends to a constant in the large-$B$ limit. In this section we comment on the role of the pion mass term for a multishell large-$B$ solution.

The simplest pion mass term that can be added to the model (\ref{L}) is
\begin{equation}
    {\cal L}_m = \frac12 m^2 f_\pi^2 \,\text{tr}\,(U-1)
    =-\frac12 m^2 \vec \pi^2+\ldots,
\end{equation}
where ellipsis stand for higher orders of pion fields. With this term the expression for the skyrmion mass (\ref{M}) becomes
\begin{equation}
    M \to M + 4 \pi m^2 f_\pi^2 \int_0^\infty (1 - 
     \cos F)r^2 dr\,,
\end{equation}
and in the r.h.s.\ of Eq.~(\ref{EOM}) one gets additional term $m^2 r^2 \sin F$. In contrast with other terms in the r.h.s.\ of this equation, we cannot discard $m^2 r^2 \sin F$ since this term grows with distance and oscillates for $F\gtrsim1$. Therefore, Eq.~(\ref{EOM-short}) gets the following correction due to the pion mass term
\begin{equation}
\label{EOM-mass}
    \left[(r^2 +kr_\text{sk}^2) F'\right]'  = m^2 r^2 \sin F\,.
\end{equation}
Unfortunately, exact analytic solutions of this equation are not known, but the large-distance asymptotics becomes $F(r)|_{r\to\infty}\propto r^{-1}e^{-mr}$. 

It would be interesting to explore other pion mass terms proposed in Ref.~\cite{Kopeliovich} and study whether either of these terms allows for explicit analytic solutions reducing to (\ref{solution}) in the massless case. We hope that such solutions may possess a constant mass per unit baryon charge in the large-$B$ limit. If such stable solutions exist, they may describe compact composite objects of baryonic matter fully within the frames of the Standard Model. Such compact composite objects could play role of dark matter particles similar to the ones studied in Refs.~\cite{QN1,QN2,QN3,QN4}. In Refs.~\cite{Gorham2012,Lawson2019} it was shown that for $B\gtrsim10^{25}$ such dark matter particles are not excluded neither by cosmological observation nor by Earth-based dark matter detectors. We leave this question for future studies.

\subsection{Towards Skyrmion stars}

In Ref.~\cite{GravSkyrm} an Einstein-Skyrme model was proposed, and in Ref.~\cite{piette2007,Nelmes2011} spherically symmetric solutions in this model were considered as models of stars similar to neutron stars. Following these works, we also consider the Skyrme model (\ref{L}) minimally interacting with a metric field, and assume a spherically symmetric ansatz for the latter:
\begin{align}
ds^2 &= -A^2(r)\left(1-\frac{2m(r)}{r} \right)dt^2
+\left(1-\frac{2m(r)}{r} \right)^{-1}dr^2\nonumber\\
&+r^2(d\theta^2 + \sin^2\theta d\varphi^2)\,.
\label{ds}
\end{align}
Here $A(r)$ and $m(r)$ are unknown functions subject to the condition that the metric (\ref{ds}) is asymptotically flat. Within this ansatz, the stationary static solutions of the Einstein-Skyrme theory obey the following equations \cite{piette2007}
\begin{subequations}
\label{system2}
\begin{align}
    \mu'&=\alpha\bigg[ \frac12 x^2 S F'^2 + \sin^2 F\left(k+k S F'^2 + \frac{{\cal I}\sin^2 F}{2x^2}\right)\bigg]\,,\\
    A'&=\alpha \left( x+\frac{2N}{x}\sin^2 F \right)AF'^2\,,\\
    &\left[(x^2 + 2k\sin^2 F)ASF'\right]' = A\sin(2F)
    \bigg( k +kSF'^2\nonumber \\& +\frac{{\cal I}\sin^2 F}{x^2}\bigg)\,,
\end{align}
\end{subequations}
where $x \equiv \frac{r}{r_\text{sk}}=ef_\pi r$, $\mu = ef_\pi m(r)$, and $S = 1-\frac{2m}{r} = 1-\frac{2\mu}{x}$ are dimensionless variables and 
\begin{equation}
\alpha = 4\pi G f_\pi^2 = 7.3\times 10^{-40}
\end{equation}
is a dimensionless coupling constant. The derivatives in Eqs.~(\ref{system2}) are over the dimensionless variable $x$.

When the function $F(x)$ is large, $F\gg1$, the average values (\ref{average}) of the trigonometric functions in Eqs.~(\ref{system2}) can be assumed. Thus, for a multishell solution with $b\gg1$, Eqs.~(\ref{system2}) simplify drasically:
\begin{subequations}
\label{system3}
\begin{align}
    \mu'&=\frac{\alpha}2\bigg[ (x^2+k) S F'^2 + k + \frac{3{\cal I}}{8x^2}\bigg]\,,\label{system3a}\\
    A'&=\alpha \frac{ x^2+k}{x} AF'^2\,,\label{system3b}\\
    &\left[(x^2 + k)ASF'\right]' = 0\,.\label{system3c}
\end{align}
\end{subequations}
The latter equation may be integrated once, and the remaining equations may be treated perturbatively with the small parameter $\alpha$. 

Although the application of Eqs.~(\ref{average}) provides a drastic simplification of equations in the Einstein-Skyrme model, this model in the regime of weak gravitational field still suffers from instability against decays into single-shell or single-skyrmion configurations. We hope that this problem will be resolved elsewhere.

\vspace{3mm}
\textit{Acknowledgements} --- IBS is grateful to Olaf Lechtenfeld for useful comments.
The work was supported by the Australian Research Council Grants No.\ DP230101058 and DP200100150.

%%%%%%%%%%%%%%%%%%%%%%%%%%%%%%%%%%%%%%%%%%%%%%%%%%%%%%%%%%%%%%%%%%%


\end{document}